# Unusual Interactions of Pre-and-Post-Selected Particles

Y. Aharonov[1,2], E.Cohen[1], S.Ben-Moshe[1]

[1]School of Physics and Astronomy, Tel-Aviv University, Tel-Aviv 69978, Israel
[2]Schmid College of Science, Chapman University, Orange, CA 92866, USA

**Abstract.** Weak value is increasingly acknowledged as an important research tool for probing quantum pre- and post-selected ensembles, where some extraordinary phenomena occur. We generalize this concept to the broader notion of "weak potential" which enables predicting the interactions between particles when one of them is pre-/post-selected and the interaction potential is small. A harmonic oscillator is considered, undergoing weak position and momentum measurements between strong position measurement, and shown to possess peculiar physical properties, affecting the momentum rather than the position of another oscillator interacting with it.

## 1 Introduction

In classical physics, pre- and post-selection have a simple, actually trivial meaning. An ensemble of $N$ identical objects undergoes a measurement at time $t_i$. All $n_i$ objects found to possess physical value $p$ are selected. Then, at time $t_f$, this $n_i$ sub-ensemble undergoes another physical measurement, and those $n_f$ objects found to possess physical value $q$ are selected again. At any intermediate time $t_i < t < t_f$, each object within the sub-sub-ensemble $n_f$ possesses both $p$ and $q$. Equally trivial is the causal account of this result: All $n_f$ objects *have possessed* these two values *all along*, with the two selections doing just that, namely, selecting them out of all others.

Things are radically different in quantum mechanics, where measurement does not detect a pre-existing value but actively *realizes* it out of several, possible ones that potentially coexist (superposition), turning that value into the only existing one ("collapse"). Therefore, in the above case, $p$ becomes real only after the first measurement, and $q$ after the second. However, should the two measurements be noncommuting, such as position and momentum, then the second measurement would blur out the first, such that, at the end, all selected objects have definite $q$, while $p$ becomes superposed again.

In the Two-State-Vector Formalism of QM (TSVF) [1,2], the account is again markedly different. Measurement not only actively determines the measured value, but it does that *for an entire segment of its history, extending to the next measurement, in both time directions*. Therefore, in the above case, the physical parameters of all objects in the sub-sub-ensemble are consistent with both values *during the entire $(t_i, t_f)$ interval*.

The last statement seems to run counter to the uncertainty principle, which forbids noncommuting variables to have precise values at the same time. Yet, another product of the TSVF, namely, weak measurement [3], has validated this very prediction in numerous experiments [4,5].

Once, therefore, this unique state proves to be valid, it should not be surprising that other physical properties of it are equally unusual [6,7,8].

In this article we point out a new physical oddity exhibited by an ensemble of almost classical particles that undergo pre-and post-selection. We show that such a particle's interaction with another particle ends up with the latter changing not its position, as would be expected, but its momentum. In order to do so, we shall need first to develop the mathematical formulation of "weak potential" which is a generalization of the weak value [2].

## 2 The Weak Potential

We shall present now the concept of weak potential using a general two-particles Hamiltonian with a small interaction term.
Let the Hamiltonian

$$H(1,2) = H_1(1) + H_2(2) + \lambda V(1,2) \equiv H_0(1,2) + \lambda V(1,2) \tag{1}$$

govern a two-body system, where $H_i(i)$ is a one-body Hamiltonian effecting particle $i$, $V(1,2)$ is a two-body interaction and $\lambda \ll 1$ is a small parameter. Let the system be described in Schrödinger picture where the time evolution of the system is $\Psi(1,2,t)$. One particle is pre- and post-selected with $|\psi_1\rangle$ at $t=0$, $\langle\psi_2|$ at $t=\tau$, and the second particle is only pre-selected with $|\phi\rangle$ at $t=0$. Let the time interval $[0, \tau]$ be divided into $N$ equal $\Delta t$ intervals.
The time evolution of the initial state within first order of perturbation theory[2] is:

$$|\Psi(1,2,t)\rangle \approx |\psi_1(1,\tau)\phi(2,\tau)\rangle = U(\tau)|\psi_1(1,0)\phi(2,0)\rangle \approx$$
$$\approx e^{-i\int H dt/\hbar}|\psi_1(1,0)\phi(2,0)\rangle \tag{2}$$

Using the interval division, we omit, to first order in λ, the non-commuting terms, and get:

$$\approx T \underbrace{e^{-iH\Delta t/\hbar} \cdot e^{-iH\Delta t/\hbar} \cdot \ldots \cdot e^{-iH\Delta t/\hbar}}_{N \ times}|\psi_1(1,0)\phi(2,0)\rangle \approx \tag{3}$$
$$\approx T \underbrace{e^{-i\lambda V\Delta t/\hbar} \cdot \ldots \cdot e^{-i\lambda V\Delta t/\hbar}}_{N \ times} \underbrace{e^{-iH_2\Delta t/\hbar} \cdot \ldots \cdot e^{-iH_2\Delta t/\hbar}}_{N \ times} e^{-iH_1\tau/\hbar}|\psi_1(1,0)\phi(2,0)\rangle \approx$$
$$= T e^{-i\lambda V\tau/\hbar} e^{-iH_2\tau/\hbar}|\psi_1(1,\tau)\phi(2,0)\rangle,$$

Where $T$ is the time order operator.
Multiplying (3) by the post-selected state $\langle\psi_2(1,\tau)|$ gives:

$$\langle\psi_2(1,\tau)|\psi_1(1,\tau)\rangle|\phi(2,\tau)\rangle = T\langle\psi_2(1,\tau)|e^{-i\lambda V\tau/\hbar}|\psi_1(1,\tau)\rangle)e^{-iH_2\tau/\hbar}|\phi(2,0)\rangle \tag{4}$$

or:

$$|\phi(2,\tau)\rangle = Te^{-i[H_2+V_w(\tau)]\tau/\hbar}|\phi(2,0)\rangle + O(\lambda^2) \tag{5}$$

where

---

[2] We assume that: $\psi_i = \psi_i^0 + \lambda\psi_i^1 + \lambda^2\psi_i^2 + \ldots$, so: $\psi_1\psi_2 = \psi_1^0\psi_2^0 + O(\lambda)$.

$$V_w(\tau) = \frac{\langle \psi_2(1,\tau)|V|\psi_1(1,\tau)\rangle}{\langle \psi_2(1,\tau)|\psi_1(1,\tau)\rangle} \qquad (6)$$

is the "weak potential".

In higher orders, changes in $|\psi_1\rangle$ and $\langle\psi_2|$ due to the interaction potential should be taken into consideration.

To second order, the effective potential would be:

$$V_W(2) = \lambda \frac{\langle \psi_2^0|V|\psi_1^0\rangle}{\langle \psi_2|\psi_1\rangle} + \frac{\lambda^2}{2}[\frac{\langle \psi_2^0|V|\psi_1^1\rangle}{\langle \psi_2|\psi_1\rangle} + \frac{\langle \psi_2^1|V|\psi_1^0\rangle}{\langle \psi_2|\psi_1\rangle} + \frac{\langle \psi_2^0|V^2|\psi_1^0\rangle}{\langle \psi_2|\psi_1\rangle}] \qquad (7)$$

where $V^2$ is the second-order correction to the potential. But this effective potential depends on the effect of the perturbation: backward/forward/both. We expect to encounter the second component in the backward case and the third in the forward one.

As an example, consider the following Hamiltonian

$$H = \frac{p_1^2}{2m} + \frac{p_2^2}{2m} + \frac{1}{2}m\omega^2 x_1^2 + \frac{1}{2}m\omega^2 x_2^2 + \lambda x_1 x_2, \quad \lambda \ll 1 \qquad (8)$$

describing two coupled harmonic oscillators with the same mass and frequency. The pre-/post-selection states are chosen to be:

$$|\psi_1\rangle = |0\rangle - i|1\rangle + |2\rangle, \quad |\psi_2\rangle = |0\rangle + |1\rangle - |2\rangle \qquad (9)$$

where $|n\rangle$ describes the $n^{th}$ energy state of the unperturbed oscillators.

According to Eq. (5) the first order correction would be:

$$\lambda \frac{\langle \psi_2^0|V|\psi_1^0\rangle}{\langle \psi_2|\psi_1\rangle} = i\lambda\sqrt{\frac{\hbar}{2m\omega}}(\langle 0|+\langle 1|-\langle 2|)(a+a^\dagger)x_2(|0\rangle-i|1\rangle+|2\rangle) =$$
$$= i\lambda\sqrt{\frac{\hbar}{2m\omega}}(-i+\sqrt{2}+1+i\sqrt{2})x_2 = \lambda\sqrt{\frac{\hbar}{2m\omega}}[1-\sqrt{2}+i(1+\sqrt{2})]x_2 \qquad (10)$$

This correction would change both the momentum and position of the second particle, as opposed to the familiar interaction changing just the momentum. We discuss this phenomenon in the next section.

The main consequence is that, contrary to Parrott [9], weak values are uniquely defined when the potential is weak and pre-/post-selected states are appropriately chosen. Furthermore, they are the basic elements of a two-particle interaction and define the strength of the perturbation potential in the general case of a composite system.

We managed to perform the above calculation only by assuming that the interaction is weak. This is obviously the case when dealing with weak measurements and weak values. Parrott [9], however, omits this point when presenting the weak values, and measures the state with the strong operator B.

## 3  Peculiar Weak Values and the Correspondence Principle

To the extent that the above reasoning is sound, it is reasonable to expect that it will have a physical content detectable by experiment. What, then, is that physical content, and how can it be demonstrated?

Consider, for example, an ensemble of electrons hitting a nucleus in a particle collider. Their initial states are known, and a specific post-selection is done after the interaction. The main interaction is purely electromagnetic, but there is also a relativistic and spin-orbit correction in higher orders which can be manifested now in the form of a weak interaction.

In what follows, the weak potential will enable us to test the physical meaning of peculiar weak values possessed by a pre-/post-selected harmonic oscillator in a way that challenges the correspondence principle.

The following gedanken experiment is based on Bohr's correspondence principle, which bridges between the classical and quantum realms by showing how, the larger the object's quantum numbers, the more classical its behavior becomes. In some special cases, this principle allows an object to exhibit both quantum and classical phenomena. This is the case, e.g., with a Rydberg hydrogen atom excited such that its electron's orbit is large enough to allow the electron to behave nearly classically. Then, let a harmonic oscillator $O$ have such wide amplitude that its motion is almost classical. Next consider an ensemble of such harmonic oscillators, each described by the Hamiltonian[3] $H = (x^2 + p^2)/2$. Let each oscillator undergo two strong position measurements at $t_i$ and $t_f$, and three weak measurements at some intermediate time $t_i < t < t_f$, as follows: a position weak measurement, a momentum weak measurement, and a weak interaction with another oscillator $O'$ which serves as a test particle. We pre- and post-select the oscillators: $\psi_i = \pi^{-1/4} \exp[-(x-x_0)^2/2)]$ at $t_i$ and $\psi_f = \pi^{-1/4} \exp[-(x+x_0)^2/2)]$ at $t_f$ after an integer number of periods, where $x_0$ is a constant, satisfying $x_0 \gg 1$. This, of course, is a rare case, but with a sufficiently large initial ensemble N an appropriately large sub-ensemble $n_f$ can be selected. The measurements thus determine each particle's entire wave function: The pre- and post-selected states are two Gaussians (Fig. 1) separated by their positions. Due to the high excitation, the states obey the correspondence principle. Note that the quantum operators of position and momentum almost commute when $x_0 \gg 1$, which means they are almost classical:

$$[\frac{x}{x_0}, p] = \frac{i}{x_0} \approx 0 \tag{11}$$

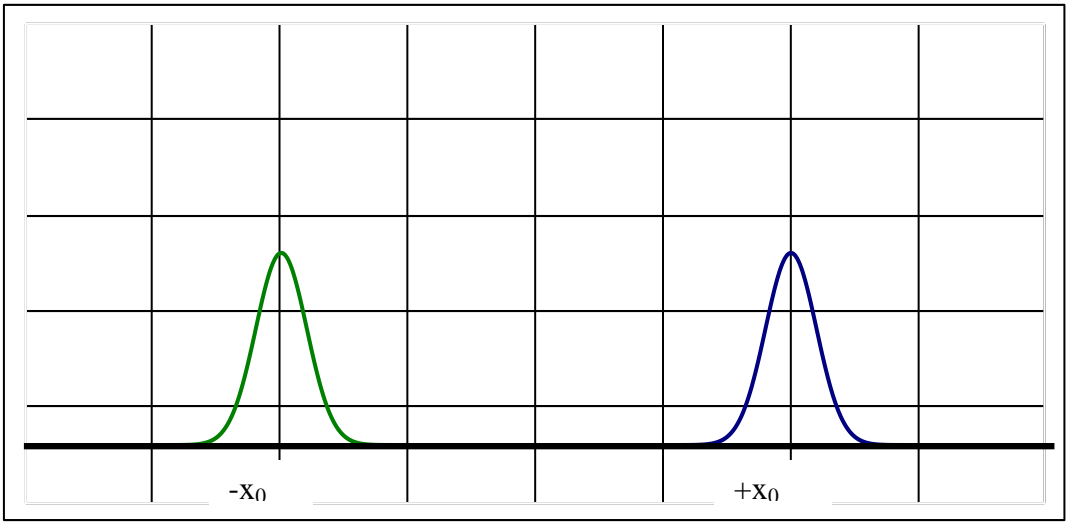

**Fig. 1.** Pre- and post-selected Gaussian states.

---

[3] Along the calculation we use nondimensionalization: Energy is measured in units of $\hbar\omega$ and distance in units of $(\hbar/m\omega)^{1/2}$.

With the aid of a weak measurement, we can find in $t=t_i$ the position's weak value:

$$x_w = 0 \tag{12}$$

*i.e.*, the particle stays precisely in the middle between its initial and final positions. More interesting is the resulting momentum weak value:

$$p_w = -ix_0 \tag{13}$$

a peculiar result in that the momentum operator's weak value turns out to be *imaginary*.
Using Eqs. (5),(6) we find for $t_i<t<t_f$:

$$x_w(t) = \frac{-ix_0}{m\omega}\sin\omega t \tag{14}$$

and

$$p_w(t) = -ix_0 \cos\omega t \tag{15}$$

In other words, in all times between $t_i$ and $t_f$, the particle's position and momentum are found to be with very high amplitude on the *imaginary* axis. While such a result is often dismissed as a calculation artifact devoid of physical content ("unphysical" [2]), it may be more rewarding to seek a situation that proves otherwise. Such a prediction is indeed offered by the weak interaction which our oscillator undergoes with a test oscillating particle $O'$ during the $t_i<t<t_f$ interval. Let $O'$ be described by a wave function $\psi_t = \exp(-p^2)$, interacting with our oscillator via an interaction term of the form, $\lambda p_1 p_2 g(t)$ where $p_1$ and $p_2$ are the momenta of the original and test particles, respectively, and $g(t)$ is normalized to 1. Here, it is $O'$ 's *momentum*, $p_2$, rather than its position, which changes upon weak interacting with $O$ (see Eq. 5):

$$\exp(i\int \lambda p_1 p_2 g(t)dt)\exp(-p_2^2) = \exp[\lambda x_0 \cos(\omega t)p_2]\exp(-p_2^2) =$$
$$= \exp[\lambda x_0 \cos(\omega t)p_2]\exp(-p_2^2) = \exp\{-[p_2 - \lambda x_0 \cos(\omega t)/2]^2\}\exp[\lambda^2 x_0^2 \cos^2(\omega t)/4]$$

(14)

## 4 The Significance

This phenomenon has no classical analogue. Bohr's correspondence principle demands the behavior of a system described by QM to reproduce classical physics in the limit of large quantum numbers. For large orbits and large energies, quantum calculations must agree with classical ones. This is not the case in the present experiment, where a marked deviation arises in a system which, *a priori*, should correspond only to a classical one.

As we have opted for granting Eqs. (14) and (15) physical reality, and further pointed out a case where they make experimental difference, let us more explicitly define their physical meaning. If the momentum's weak value turns out to be imaginary, then it is the energy of the original harmonic oscillator that must turn out to be negative. The bearings of allowing negative kinetic energy has already been investigated in the past [10] and shown to be consistent under the restrictions indicated therein. In the present case, however, this negative energy is not only kinetic. Rather, a harmonic oscillator's total energy, known to be bounded from below, turns out to be negative, despite the system corresponding to a classical one. Moreover, in light of the interaction of $O$ and $O'$, which is very concrete, and bearing in mind our arguments in [8], these weak values cannot be dismissed as mere errors [2] but have a real, physical existence.

Following this result, several surprising predictions of the TSVF make simple sense, such as the recently verified "Cheshire cat" effect [11]. We explore this issue in greater detail in consecutive papers [12][13].

## Acknowledgements

It is a pleasure the thank Paz Beniamini, Avshalom C. Elitzur and Doron Grossman for their valuable comments. This work has been supported in part by the Israel Science Foundation Grant No. 1125/10.

## References

1. Y. Aharonov, P.G. Bergman, J.L. Lebowitz , Phys. Rev. **134** (1964).
2. Y. Aharonov, D. Rohrlich, *Quantum Paradoxes: Quantum Theory for the Preplexed*, Wiley, Weinheim (2005).
3. Y. Aharonov, D. Z. Albert, L. Vaidman *Phys. Rev. Lett* **60**, 1351-1354 (1988),.
4. J. S. Lundeen, B. Sutherland, A. Patel, C. Stewart, C. Bamber, Nature **474**, 188–191 (2011).
5. S. Kocsis, B. Braverman, S. Ravets, M.J. Stevens, R.P. Mirin, L.K. Shalm, A.M. Steinberg, Science **332**, 1170 (2011).
6. Y. Aharonov, E. Cohen, A. C. Elitzur, Phys. Rev. A (submitted). http://arxiv.org/abs/1207.0655.
7. Y. Aharonov, E. Cohen, A. C. Elitzur, Phys. Rev. A (submitted). http://arxiv.org/abs/1207.0667.
8. Y. Aharonov, E. Cohen, D. Grossman, A. C. Elitzur, *International Conference on New Frontiers in Physics* (to be published). http://arxiv.org/abs/1206.6224.
9. S. Parrott, http://arxiv.org/pdf/0909.0295v3.pdf (2011).
10. Y. Aharonov, S. Popescu, D. Rohrlich, L. Vaidman, Jap. Jour. App. Phys. **9**, 251-253 (1993).
11. Y. Aharonov, S. Popescu, P. Skrzypczyk, Quantum Cheshire Cats. http://arxiv.org/pdf/1202.0631.pdf (2012).
12. Y. Aharonov, E. Cohen, A. C. Elitzur, PRL, (submitted).
13. Y. Aharonov, E. Cohen, A. C. Elitzur, PRL (submitted).